\documentclass[conference]{IEEEtran}
\usepackage{todonotes}
\usepackage{amsmath,amssymb,amsfonts}
\DeclareMathOperator*{\argmax}{arg\,max} 
\DeclareMathOperator*{\argmin}{arg\,min}
\usepackage{algorithmic}
\usepackage{graphicx}
\usepackage{textcomp}
\usepackage{xcolor}
\usepackage{braket}
\usepackage{float}
\usepackage{flushend}
\usepackage[caption=false]{subfig}
\usepackage{graphicx}
\usepackage[utf8]{inputenc}
\usepackage[T1]{fontenc}
\usepackage{bm}

\usepackage{listings}
\usepackage{color}
\lstdefinestyle{mystyle}{
    backgroundcolor=\color{backcolour},   
    commentstyle=\color{codegreen},
    keywordstyle=\color{magenta},
    numberstyle=\tiny\color{codegray},
    stringstyle=\color{codepurple},
    breakatwhitespace=false,         
    breaklines=true,                 
    captionpos=b,                    
    keepspaces=true,                 
    numbers=left,                    
    numbersep=5pt,                  
    showspaces=false,                
    showstringspaces=false,
    showtabs=false,                  
    tabsize=2
}

\def\BibTeX{{\rm B\kern-.05em{\sc i\kern-.025em b}\kern-.08em
    T\kern-.1667em\lower.7ex\hbox{E}\kern-.125emX}}

\usepackage{cite}

\usepackage{hyperref}    
\begin{document}
\bstctlcite{IEEEexample:BSTcontrol}
\title{Towards a Hybrid Quantum Enhanced Solution for Densest $k$-Subgraph Problem}

\author{\IEEEauthorblockN{Ravi Sangwan} \IEEEauthorblockA{\textit{TCS Research} \\ \textit{Tata Consultancy Services Ltd}\\ India \\ ravi.sangwan@tcs.com} \and \IEEEauthorblockN{Prabhat Anand} \IEEEauthorblockA{\textit{TCS Research} \\ \textit{Tata Consultancy Services Ltd}\\ India \\ anand.prabhat@tcs.com} \and \IEEEauthorblockN{M Girish Chandra} \IEEEauthorblockA{\textit{TCS Research} \\ \textit{Tata Consultancy Services Ltd}\\ India \\ m.gchandra@tcs.com}}

\maketitle
\begin{abstract}
Gaussian Boson Sampling (GBS) excels as a native quantum primitive for the unconstrained Dense Subgraph Problem (DSP). However, adapting GBS to the Densest k-Subgraph Problem (DkSP) causes sample inefficiencies due to hard post-selection, especially under realistic photon-loss conditions. We overcome this bottleneck using a hybrid quantum-classical framework with lightweight heuristic-based refinement strategies to transform near-k samples into valid k-subgraph solutions. Using numerical simulations across synthetic and several standard empirical networks such as Zachary’s Karate Club, the proposed methods improve solution density relative to hard post-selection and outperform classical baselines on sparse random graphs. They also achieve a $4$ × improvement in sample efficiency in clustered graphs and maintain robustness under realistic photon-loss conditions. A quality-normalized resource analysis, combining analytical operation counts with measured classical processing times, further quantifies the trade-off between quantum-sample efficiency, classical computational cost, and solution quality. Overall, pairing noisy GBS sampling with lightweight classical projection offers a practical, resilient hybrid framework for combinatorial graph optimization.\end{abstract}
\section{Introduction}
The Densest $k$-Subgraph Problem (D$k$SP) is a central combinatorial optimization task in graph theory with widespread applications in domains like network science and data mining \cite{DkS2,DSD_Tutorial}. The importance of D$k$SP lies in its ability to capture latent patterns of relationships that often correspond to meaningful structural or functional units in complex systems. For example, dense subgraphs within social networks can be interpreted as communities, cliques, or groups of individuals with strong mutual interactions \cite{DSP_Community_detection, Connectivity_Social_network}. Dense substructures found in the case of financial transaction networks may indicate coordinated fraud, and in communication or cyber-security domain, these substructures indicate anomalous clusters of traffic or coordinated attacks \cite{DSP_Application, spotlight_graph}.   

It is a constrained Densest Subgraph Problem (DSP) where the objective is to find a subgraph with $k$ vertices (nodes) with highest density. D$k$SP is comparatively hard for sparse graphs in comparison to dense graphs pertaining to their high probability of a good guess. DkSP is NP-hard and is widely regarded as difficult to approximate in contrast to the unconstrained densest subgraph problem, no general polynomial-time constant-factor approximation algorithm is known for DkSP~\cite{DkS1}. Compared to the unconstrained DSP, which can be solved in polynomial time by flow-based algorithms, placing a size-constricted solution has a strong impact on computational complexity~\cite{flow_based_DSP}. D$k$SP generalizes the Maximum Clique Problem (MCP) known over many years~\cite{MCP}. If the maximum possible density is one in case of D$k$SP, an optimal solution is an exact clique over a size of $k$.

From a complexity‑theoretic perspective, the decision version of D$k$SP entails a combinatorial solution space of size $\Theta(\binom{n}{k})$, making brute‑force enumeration intractable under classical complexity models. 
Unlike classical Turing or von Neumann machines, quantum computation operates as a probabilistic linear‑algebraic model, in which exponentially large combinatorial spaces can be coherently represented within a compact Hilbert space. Quantum algorithms do not aim to deterministically identify optimal solution, but instead bias measurement outcomes toward high‑quality solutions. This naturally leads to quantum sampling formulations of D$k$SP, among which Gaussian Boson Sampling (GBS) is particularly relevant, as its output statistics are intrinsically tied to subgraph densities and is natively realizable on photonic quantum architectures.

GBS provides a non-universal quantum computer model, employing squeezed light, linear interferometers and photon detectors \cite{gbs}. It has been shown that GBS is biased towards producing samples with high density \cite{gbs_dsp}. This bias is due to the correlation between the probabilities of detecting photons and the hafnians of submatrices associated with perfect matchings of underlying graphs, which are themselves highly correlated with subgraph edge density \cite{Locally_DSP}. Subsequently, demonstrations of D$k$SP using GBS have been reported in graph pattern discovery, such as community detection, molecular docking, and data mining \cite{Graph_problem_GBS, Mol_docking_GBS}. Additionally, D$k$SP can be naturally used  to test the possibility of any quantum advantage with current photonic architectures \cite{DSP_GBS_Exp}.

However, realizing a practical computational advantage of GBS for D$k$SP requires addressing a key bottleneck. 
\textit{Sampler efficiency:} In the post-selection strategy, only the GBS samples with exactly $k$ detected modes are retained, while discarding potentially high-quality samples whose sizes are not equal to $k$. 

To address this limitation we propose three post processing techniques that utilize all potentially good samples whose sizes are slightly smaller or larger than $k$, i.e., within a small deviation relative to \(k\), e.g., \(|S|\in[k-2,k+2]\) for \(k=10\), and convert them into valid $k$‑vertex solutions while aims to preserve or improve edge connectivity: (i) degree‑based greedy resizing, (ii) edge‑based greedy resizing, and (iii) an ACO‑based meta‑heuristic repair and refinement procedure. These techniques enable the exploitation of more GBS samples, including infeasible ones, by deterministically or heuristically adding or removing vertices to improve the edge-connectivity and \textit{sampler efficiency}. 
 
For a comprehensive evaluation, GBS with post-selection and processing techniques were compared against a set of classical methods, including uniform sampling, \emph{degree‑based} and \emph{edge‑based} classical samplers, the deterministic \emph{Charikar's} algorithm \cite{charikar_dsp}, Ant Colony Optimization (\emph{ACO}) \cite{ACO_DkSP}, and \emph{brute‑force} search as an accuracy benchmark for small graph instances. Extensive simulations on random graphs demonstrate that edge‑based and ACO‑based GBS post‑processing consistently achieve improves densities with significantly fewer samples than GBS with post-selection and classical sampling methods, while exhibiting improved convergence and \textit{sampler efficiency}.

The rest of the paper is organized as follows. Section~\ref{sec:Dksp} provides the necessary background on the Densest $k$-Subgraph Problem (D$k$SP), classical solution approaches, and Gaussian Boson Sampling for D$k$SP. Section~\ref{sec:methods} presents the proposed hybrid quantum-classical framework, including the GBS post-processing strategies and the associated cost analysis. Finally, Section~\ref{sec:simulations_results} presents the numerical results and comparative performance evaluation.
\section{Background}
\subsection{Densest k-Subgraph Problem}\label{sec:Dksp}
Let $G = (V,E)$ be an unweighted, undirected graph with vertex set $V$, edge set $E$, and total number of vertices (nodes)  $|V| = n$. This graph is represented by its adjacency matrix $A \in \{0,1\}^{n \times n}$, where $A_{ij} = 1$ if and only if $(i,j) \in E$, i.e., $i^{th}$ abd $j^{th}$ vertices are connected with an edge and $A_{ij} = 0$ otherwise. For any subset of vertices $S \subseteq V$ with fixed cardinality $|S| = k$ (i.e., total of $k$ vertices), let $G[S]$ denote the subgraph induced by $S$, and let $E(S)$ denote the set of edges where both endpoints (vertices) lie in $S$. The number of induced edges can be written equivalently as
\begin{equation}\label{eq:no_edges}
    |E(S)| = \frac{1}{2} \sum_{i,j \in S} A_{ij}.
\end{equation}
The density ($D$) of the induced subgraph $G[S]$ is defined as the ratio of number of edges in subgraph to the total maximum possible number of edges  
\begin{equation}\label{eq:normal_density}
D(S) = \frac{2|E(S)|}{k(k-1)}.
\end{equation} Note that a $k$-vertex subgraph can contain at most $\binom{k}{2} = \frac{k(k-1)}{2}$ edges. We can also define density as the ratio of the number of edges to the number of vertices but to keep the density normalized, we will follow \eqref{eq:normal_density} in this work. The density ($D$) in \eqref{eq:normal_density} satisfies $0 \le D(S) \le 1$, with $D(S)=1$, iff $G[S]$ is a clique and $D(S)=0$ if all the vertices in $S$ are mutually disconnected. The D$k$SP seeks a vertex subset, $S^\ast$ (the optimal subset) that maximizes the induced density over all subsets of fixed size $k$
\begin{equation}
S^\ast = \argmax_{S \subseteq V,\ |S| = k} D(S).
\end{equation}
Although it seems to be formulated quite simply, D$k$SP summarizes the basic computational problem, to identify very strong cohesive units hidden in large-scaled networks.

The D$k$SP has been attempted by a range of classical methods including exact algorithms, approximation algorithms, samplers and meta-heuristic methods~\cite{Survey_DSP_Variants,DkSP_large_graphs}. Here, we discuss a few of the algorithms that promise superior trade-off between efficiency and accuracy and further try to place GBS on this spectrum.

\subsection{Classical Approaches}\label{sec:classical methods}
\subsubsection{Brute-force Approach}
In \emph{brute-force} algorithm, one lists all the possible $k$-vertex subsets $\binom{n}{k}$ and counts the number of edges for all subsets in order to calculate their densities and return the $k$-vertex subset with maximum density. It ensures optimality, but evaluates all $\binom{n}{k}$ subsets, which makes the method infeasible for larger graphs as $n$ and $k$ increases.  

\subsubsection{Charikar's Algorithm}
The most prominent deterministic approach for DSP is \emph{Charikar's Algorithm}, which is the greedy peeling algorithm for the DSP in which we start with full graph, remove the minimum-degree vertices until only one vertex remains and obtain the subgraph with maximum density, where the degree of a vertex is the total number of nodes connected to it \cite{charikar_dsp}. \emph{Charikar's Algorithm} can be modified to D$k$SP by scanning intermediate subgraph of size $k$ obtained during the process of repeated minimum-degree node removal \emph{Charikar's Algorithm} does not guarantee optimality because it gives the subset of the $k$ vertices with relatively high degree and that doesn't guaranteed that these high degree vertices are well connected among themselves \cite{Finding_DS}. As a result, the obtained solution may not correspond to the densest possible $k$-subgraph. This shortcoming motivates the exploration of more adaptive search strategies that better capture the internal connectivity among selected vertices.

\subsubsection{Ant Colony Optimization}
Besides deterministic approximation algorithms, meta-heuristic methods have been shown to be quite effective even in large-scale graphs. \emph{Ant Colony Optimization (ACO)} is considered a powerful meta-heuristic for combinatorial optimization problems, including subgraph selection~\cite{ACO,ACO_DkSP}.  In the \emph{ACO} based D$k$SP algorithm, each artificial ant starts with selecting an initial vertex (or set of vertices) either randomly or based on other heuristic techniques (such as degree-based), and incrementally  constructs  a  subset  of $k$ vertices by choosing a candidate vertex based on a selection score that  combines  pheromone value  and  the heuristic gain ($g$). Initially, the pheromone values for all vertices are equal. At each iteration, all ants construct candidate subgraph solutions, and the pheromone values are updated through persistence and reinforcement. Specifically,
\begin{equation}
p(v) \leftarrow \rho \, p(v),
\label{eq:aco_evaporation}
\end{equation}
where $p(v)$ is the pheromone value associated with vertex $v$ and $\rho \in (0,1)$ is the persistence factor. This is followed by reinforcement,
\begin{equation}
p(v) \leftarrow p(v) + \sum_{(S,d)\in \mathcal{E}} \mathbf{1}_{\{v \in S\}} \, d,
\label{eq:aco_reinforcement}\end{equation}
where $\mathcal{E}$ denotes the set of elite solutions (top-ranked solutions by density), $S$ is a selected $k$-vertex subset, $d$ is its density, and $\mathbf{1}_{\{v \in S\}}$ is an indicator function. This update favors vertices appearing in high-density subgraphs while maintaining exploration through persistence. The heuristic gain measures the number of new edges a candidate vertex contributes to the partially constructed subgraph. One way to define the score ($s$) for a vertex $v$ is: 
\begin{equation}\label{eq:aco_score}
\begin{aligned}
s(v) &= p(v)\cdot \big(\mathrm{g}(v)+1\big)^2, \\
\mathrm{g}(v) &= \left|\{u\in S^{(t)} : (u,v)\in E\}\right|,
\end{aligned}
\end{equation}
where $p(v)$ denotes  the  pheromone  value  associated to $v$ and $S^{t}$ is currently construed subset at step $t$. The vertex selection is probabilistic, where vertices with a high score are more likely to be chosen. \emph{ACO} provides a balance between scalability and solution quality, making it well suited for large-scale combinatorial optimization problems.

In addition to deterministic and meta-heuristic algorithms, sampling-based approaches provide an effective framework for addressing the D$k$SP. Instead of explicitly optimizing over all possible vertex subsets, these methods repeatedly generate candidate subsets of size $k$ and retain the densest solution observed. Sampling-based techniques are particularly relevant to our comparison study with GBS, since GBS itself produces candidate subgraphs through a probabilistic sampling process. 

\subsubsection{Uniform Sampling}
The simplest classical sampling baseline is \emph{uniform sampling}, in which subsets of size $k$ are drawn uniformly at random from the vertex set. Formally, every subset $S \subseteq V$ with $|S| = k$ is selected with equal probability:
\begin{equation}
P(S) = \frac{1}{\binom{n}{k}}.
\end{equation}
Each sampled subset is evaluated using the induced density, and the best solution observed after a fixed number of samples is recorded. Although this method is straightforward to implement, it explores the exponentially large search space without exploiting any structural information about the graph.

\subsubsection{Degree-based sampling}
To incorporate structural information, we can consider \emph{degree-based sampling}. In this approach, vertices are selected with a probability proportional to their degrees. The probability of selecting a vertex $v \in V$ is therefore given by
\begin{equation}
P(v) = \frac{d(v)}{\sum_{u \in V} d(u)}.
\end{equation}
A subset of size $k$ is then constructed by sampling vertices without replacement according to this distribution and after each selection, probabilities are renormalized. Since dense subgraphs typically contain vertices that are highly connected within the graph, assigning a higher sampling probability to high-degree vertices increases the likelihood of selecting subsets that lie within dense regions. However, this method still treats vertices independently and does not explicitly account for pairwise edge-connectivity among the selected vertices.

\subsubsection{Edge-based Sampling}
A more structured sampling strategy is \emph{edge-based sampling}, in which a subset is constructed sequentially starting from a randomly chosen vertex. At each step, a new vertex is probabilistically selected according to its edge-connectivity to the vertices already included in the subset.
Formally, if $S$ is the current subset, the next vertex $v \notin S$ is selected with probability proportional to: 
\begin{equation} 
P(v \mid S) \propto |\{u \in S \mid (u,v) \in E\}|. 
\end{equation} 
This procedure encourages the formation of tightly connected vertex groups and increases the probability of generating subsets with a high number of induced edges.


Taken together, these classical sampling strategies provide a hierarchy of baselines with increasing structural bias. Uniform sampling explores the search space without guidance, degree-based sampling exploits local connectivity, while edge-based sampling progressively incorporates higher-order structural information. These methods therefore provide meaningful classical baselines for evaluating the performance of GBS. 

\subsection{Gaussian Boson Sampling for DkSP} \label{sec: gbs_dksp}
GBS is a photonic quantum sampling model, where Gaussian (multimode squeezed-vacuum) states interact through a linear optical interferometer and are measured with photon-number-resolving (PNR) or threshold detectors \cite{gbs}. The output state of the GBS is a cluster pattern of photons: 
\[
S = (s_1, s_2, \dots, s_m),
\]
where $s_i$ is the count of photons detected in mode $i$, $N$ represents the total number of photons, i.e. $N = \sum_{i=1}^m s_i$ and $m$ is total number of modes present.

To employee GBS for graph problems, every graph vertex  is encoded as an optical mode (i.e., $m = |V|$). In threshold GBS, each mode produces a binary outcome: $s_i=0$ indicates no photon detection, while $s_i=1$ indicates one or more detected photons. Therefore a click in  the $i$th mode signifies the presence of $i$th vertex in the sampled subgraph. Hence every photon pattern of detection is naturally associated with a vertex subset $S \subseteq V$, whose size is equal to the total number of modes in which photons are detected.
The probability of detecting a pattern $S$ is given by:
\begin{equation}\label{eq:prob_GBS}
P(S) = \frac{|\mathrm{Haf}(A_S)|^2}{s_1! \cdots s_m! \, \sqrt{\det(\sigma_Q)}},
\end{equation}
where $\sigma_Q = \sigma + I/2$, $\sigma$ is the covariance matrix of the input Gaussian state, $I$ is identity matrix, and $A_S$ is the principal submatrix of $A$ corresponding to the sampled vertex set.
The hafnian $\mathrm{Haf}(A_S)$, is a combinatorial quantity which in case of adjacency-structured matrices is proportional to the number of perfect matchings of the graph or equal to number of perfect  matching for the case in which elements of adjacency matrices are $0$ and $1$ ( i.e., for unweighted  and undirected graphs). A perfect matching in a graph is a set of edges such that each vertex in the graph is incident to exactly one selected edge, i.e., the vertices are paired without overlap. A higher number of edges generally increases the number of possible perfect matchings, reflecting a richer connectivity structure. 
The application of GBS in graph theory is based on the combinatorial interpretation of the hafnian~\eqref{eq:prob_GBS}. As the level of connectivity increases, the number of perfect matchings tends to rise. In other words, the subgraphs with richer connectivity generally exhibit larger hafnian values. Consequently, the GBS distribution satisfies
\begin{equation}
P(S) \propto |\mathrm{Haf}(A_S)|^2,
\end{equation}
which signifies that the sampling is biased toward subsets with high structural connectivity, making it a suitable candidate for dense subgraph discovery. Unlike classical uniform sampling, GBS captures higher-order correlations between vertices. 

The undirected, unweighted graph $G= ( V,E)$  with adjacency matrix $A$ is  embedded in GBS via a rescaled matrix~\cite{Effect_of_errors_QEDSP}.
\begin{equation}
\tilde{A} = cA, \qquad c < \frac{1}{\lambda_{\max}},
\end{equation}
where $\lambda_{\max}$ is the largest eigenvalue of $A$, ensuring physical realizability of the Gaussian state. Under this encoding, the submatrix $A_S$ corresponds to the induced subgraph on the sampled vertex set \cite{Applications_PQC}.

Nonetheless, a key limitation of GBS for D$k$SP that the number of modes in which photons are detected varies across samples. As a result the subsets sampled may not meet the fixed cardinality constraint ($|S|=k$). In the post-selection strategy, only samples with exactly $k$ detected modes are retained, this approach is often insufficient, because it fails to explore the potentially dense samples whose sizes are close to, but not exactly, $k$. 


\section{Methods}\label{sec:methods} 
We propose a hybrid quantum-classical framework for D$k$SP in which GBS provides structure-biased candidate subgraphs and classical post-processing enforces the fixed-cardinality constraint. Instead of discarding all GBS outcomes whose cardinality differs from $k$, the proposed framework admits near-$k$ samples and refines them into feasible $k$-vertex solutions. We first describe the proposed post-processing strategies and then introduce a resource model to quantify the additional classical cost relative to the quantum sampling resources required to achieve a prescribed solution quality.

\subsection{GBS Post-Processing}\label{sec:post_processing_gbs}
The proposed post-processing strategies adjust the sampled subset by iteratively adding or removing vertices until the cardinality constraint is satisfied. In the degree-based method, vertices are selected or removed according to their degree, whereas in the edge-based method, the refinement is based on the number of edges a vertex contributes to the current subset. The ACO-based method uses a meta-heuristic refinement procedure to guide the selection of vertices.

For an undirected graph $G=(V,E)$ with $n=|V|$, let $\tilde{S}\subseteq V$ denote the vertex subset obtained from a threshold GBS sample. The objective is to construct a valid subset $S\subseteq V$ satisfying $|S|=k$, while maximizing the density of the induced subgraph. To achieve this, we iteratively add vertices to $\tilde{S}$ when $|\tilde{S}|<k$ and remove vertices when $|\tilde{S}|>k$, until the desired cardinality is obtained.

Related ideas have appeared in prior work on GBS-based clique finding~\cite{Mol_docking_GBS}. In that work, two post-processing techniques were introduced to convert sampled subgraphs into valid cliques (where clique is the subgraph with unit density). The first is greedy shrinking, in which low degree vertices are removed iteratively from GBS sampled subgraphs until a clique is left behind. The second is local search which starts from a valid clique ( obtain  either from GBS samples or after greedy shrinking) and proceeds in two stages. First is grow stage, where vertices which are connected to all the vertices in the current clique are identified and then added to obtain the larger clique. If no such vertices exist, it  moves to swap stage. Here, a vertices which are connected to all the vertices except one in the current clique are identified and replaced with the incompatible one in the clique, to obtain more clique of the same size. 
Unlike these approaches based on iterative neighborhood exploration under unit‑density constraints, our method uses simple greedy refinement without explicit local search. Vertices are deterministically, or meta-heuristically in the ACO-based variant, added or removed based on connectivity or degree only to enforce fixed cardinality, making the approach lightweight and well suited for the D$k$SP.
\subsubsection{Degree-based post-processing}\label{sec:degree_gbs}

The first strategy is based on total number of adjacent vertices (i.e., degree). If $|\tilde{S}| < k$, vertices are added iteratively according to their degree in the original graph:
\[
v^\ast = \arg\max_{v \notin \tilde{S}} d(v),
\]
where $d(v)$ denotes the degree of vertex $v$. If $|\tilde{S}| > k$, vertices with the smallest degrees in $S$ are removed:
\[
v^\ast = \arg\min_{v \in \tilde{S}} d(v).
\]

\subsubsection{Edge-based post-processing} \label{sec: edge_gbs}

To better preserve the internal structure of the sampled subgraph, we introduce an edge-based strategy that explicitly accounts for interactions between vertices and the current subset. For a vertex $v \notin \tilde{S}$, define its connectivity to $\tilde{S}$ as
\[
\mathrm{con}\{\tilde{S}(v)\} = |\{u \in \tilde{S} : (u,v) \in E\}|.
\]
Which counts the number of edges connecting vertex $v$ to vertices already present in the current subset $S$. 
If $|\tilde{S}| < k$, the vertex with maximum connectivity to the current subset is added:
\[
v^\ast = \argmax_{v \notin \tilde{S}}\mathrm{con}\{\tilde{S}(v)\}.
\]
If $|\tilde{S}| > k$, the vertex with minimum connectivity to the remainder of the subset is removed:
\[
v^\ast = \argmin_{v \in \tilde{S}} \mathrm{con}{\tilde{S}(v)} .
\]

The edge-based strategy directly promotes the preservation of edges within the subset and is therefore more closely aligned with the objective of maximizing induced density. In contrast, the degree-based method provides a computationally efficient baseline that relies solely on global structural properties of the graph.

\subsubsection{ACO-Based Post-Processing} \label{sec:gbs_aco_post} In addition to the proposed greedy-based post-processing techniques, we propose a metaheuristic post-processing that uses ACO as a repair and refinement procedure for infeasible GBS samples, i.e., samples satisfying $|\widetilde{S}|\neq k$. For $|\widetilde{S}|=k$, we directly evaluate the sample density. For samples with $|\widetilde{S}|\neq k$, we run the ACO routine initialized around $\widetilde{S}$ to obtain a feasible subset $S$ satisfying $|S|=k$. To keep the search localized around the GBS-generated candidate, each ant is initialized with information from $\widetilde{S}$. During each ACO iteration, each ant independently constructs a complete repaired candidate by adding vertices when the current subset is undersized and removing vertices when it is oversized. The selection of each vertex is guided by the pheromone-based selection score defined in Eq.~\eqref{eq:aco_score}. In this post-processing setting, the heuristic gain $g(v)$ measures the number of edges that a candidate vertex contributes to the current subset $S^{(t)}$. For removal, vertices with smaller internal edge contributions are preferred. The pheromone values are initialized uniformly and boosted for vertices belonging to the GBS sample: \begin{equation} \tau(v) = \begin{cases} \tau_0+\delta, & v\in\widetilde{S},\\ \tau_0, & v\notin\widetilde{S}, \end{cases} \label{eq:gbs_aco_init} \end{equation} where $\tau_0$ is the initial pheromone value and $\delta$ controls the additional bias assigned to vertices selected by GBS. After all ants construct and evaluate feasible subsets, the pheromone values are updated using the evaporation and best-candidate reinforcement rules in Eqs.~\eqref{eq:aco_evaporation} and~\eqref{eq:aco_reinforcement}. The updated pheromone values are bounded within $[\tau_{\min},\tau_{\max}]$ to reduce premature convergence. This procedure is repeated for at most $I$ iterations, with $a$ ants constructing repaired candidates in each iteration. Early termination may be applied when the target solution quality is attained or when the best density does not improve over a prescribed number of consecutive iterations. The best density obtained across all ants, iterations, and repaired GBS samples is reported. 

Unlike greedy resizing, which deterministically adds or removes vertices based on a single criterion, the ACO-based strategy explores multiple candidate repairs through pheromone-guided sampling. Thus, it enforces the fixed-cardinality constraint while preserving the ability to search over several high-density completions or removals. 

Taken together, the proposed post-processing methods convert accepted near-$k$ GBS samples into feasible $k$-vertex solutions. Since these approaches differ in both solution quality and computational requirements, a systematic resource analysis is required.

\subsection{Cost Analysis} \label{sec: cost analysis} Existing DkSP algorithms characterize solution quality through approximation guarantees relative to the optimum. Feige et al.\ developed a polynomial-time $O(n^{\gamma})$-approximation for some $\gamma<1/3$ \cite{DkS1}, while Bhaskara et al.\ obtained an $O(n^{1/4+\eta})$-approximation in $n^{O(1/\eta)}$ time \cite{1_4_DkSP_approx}. These results illustrate the relationship between attainable solution quality and computational effort. In the hybrid GBS + classical methods, attaining a common solution quality may require different amounts of quantum and classical resources. We therefore compare the methods using a quality-normalized resource vector.

For a graph instance $g$, the relative optimality gap of method $M$ is defined as \begin{equation} \epsilon_{M,g} = \frac{D_{\mathrm{opt},g}-D_{M,g}} {D_{\mathrm{opt},g}}, \label{eq:relative_optimality_gap} \end{equation} where $D_{\mathrm{opt},g}>0$ is the exact density obtained by brute-force search and $D_{M,g}$ is the best density obtained by method $M$. Thus, $\epsilon_{M,g}=0$ indicates recovery of the optimum, while $\epsilon_{M,g}\leq0.05$ indicates that the obtained density is at least $95\%$ of the optimum. For comparison with conventional approximation guarantees, an $\alpha$-approximation for a maximization problem satisfies \begin{equation} D_{M,g} \geq \frac{D_{\mathrm{opt},g}}{\alpha}, \label{eq:alpha_approximation} \end{equation} and therefore \begin{equation} \epsilon_{M,g} \leq 1-\frac{1}{\alpha}. \label{eq:alpha_gap_relation} \end{equation} Unlike such worst-case approximation guarantees, $\epsilon_{M,g}$ is an instance-level empirical quantity evaluated relative to the exact optimum. For a prescribed target gap $\epsilon$, we define the quality-normalized resource vector as \begin{equation} \mathcal{R}_{M,\epsilon} = \left(Q_{M,\epsilon},C_{M,\epsilon}\right), \label{eq:quality_resource_vector} \end{equation} where $Q_{M,\epsilon}$ and $C_{M,\epsilon}$ denote, respectively, the quantum and classical resources consumed until method $M$ first satisfies $\epsilon_{M,g}\leq\epsilon$. For GBS-based methods, the quantum resource is quantified by the number of raw GBS samples consumed: \begin{equation} Q_{M,\epsilon} = N^{\mathrm{GBS}}_{M,\epsilon}. \label{eq:gbs_quantum_cost} \end{equation} 

For fully classical methods, \begin{equation} Q_{M,\epsilon}=0. \label{eq:zero_quantum_cost} \end{equation} 
The classical resource consumption is decomposed into the principal computational operations performed by each method: \begin{equation} \begin{aligned} C_{M,\epsilon} ={}& C_{\mathrm{pre},M,\epsilon} + C_{\mathrm{class},M,\epsilon} + C_{\mathrm{cand},M,\epsilon} \\ &+ C_{\mathrm{den},M,\epsilon} + C_{\mathrm{update},M,\epsilon}. \end{aligned} \label{eq:classical_cost_components} \end{equation} Here, $C_{\mathrm{pre},M,\epsilon}$ represents graph-dependent preprocessing, including the graph-to-Gaussian-state mapping for GBS-based methods and vertex degree computation for degree-based methods. The graph-to-Gaussian-state mapping determines the scaling parameter, squeezing parameters, and interferometer settings from the graph adjacency matrix and is performed once per graph. The term $C_{\mathrm{class},M,\epsilon}$ represents the classification of GBS samples according to the cardinality-acceptance criterion; $C_{\mathrm{cand},M,\epsilon}$ represents the operations used to construct or modify candidate subsets, including candidate scoring and vertex selection; $C_{\mathrm{den},M,\epsilon}$ corresponds to complete density evaluations; and $C_{\mathrm{update},M,\epsilon}$ denotes method-specific state-maintenance operations. Components that do not apply to a particular method are set to zero.

The accounting model records elementary algorithmic operations rather than wall-clock time. Each executed operation is assigned to exactly one cost component to avoid double counting. Each GBS outcome is returned by the sampling interface as a list of clicked-detector indices, with its cardinality directly available from the list length. Classifying the outcome as exact-$k$, accepted near-$k$, or rejected therefore requires a constant number of integer comparisons and costs $O(1)$ per sample. Consequently, the total sample-classification cost for a GBS-based method is \begin{equation} C_{\mathrm{class},M,\epsilon} = O\!\left(Q_{M,\epsilon}\right). \label{eq:sample_classification_cost} \end{equation} For fully classical methods, $C_{\mathrm{class},M,\epsilon}=0$.  

The candidate-processing cost $C_{\mathrm{cand},M,\epsilon}$ captures all operations used to construct or modify candidate subsets. These operations include accesses to already available scores, adjacency queries used to compute connectivity-based scores, score comparisons, formation and normalization of a selection distribution, deterministic or probabilistic vertex selection, executed vertex insertions or removals, and maintenance of subset membership. Let $J_{M,\epsilon}$ denote the total number of candidate-construction or modification decisions executed before the target gap is attained or the prescribed resource budget is exhausted. At decision $i$, let $m_i$ denote the number of candidate entries processed and let $\kappa_{i,M}$ denote the work required to process one candidate entry. The total candidate-processing cost is therefore \begin{equation} C_{\mathrm{cand},M,\epsilon} = O\!\left( \sum_{i=1}^{J_{M,\epsilon}} m_i\kappa_{i,M} \right). \label{eq:candidate_processing_cost} \end{equation} Here, $\kappa_{i,M}=O(1)$ when candidate selection uses already available scores, whereas $\kappa_{i,M}=O(|S_i|)$ when each candidate is scored according to its connectivity with the current subset $S_i$. For methods with $J_{M,\epsilon}=0$, the candidate-processing cost is set to $C_{\mathrm{cand},M,\epsilon}=0$.

For a complete candidate subset $S$ satisfying $|S|=k$, evaluating its density requires checking every unordered pair of distinct vertices. The exact cost of one density evaluation is \begin{equation} c_{\mathrm{den}}(k) = \binom{k}{2} = \frac{k(k-1)}{2}. \label{eq:single_density_cost} \end{equation} Consequently, if $N_{\mathrm{den},M,\epsilon}$ complete candidates are evaluated, the exact density-evaluation cost is \begin{equation} C_{\mathrm{den},M,\epsilon} = N_{\mathrm{den},M,\epsilon} \binom{k}{2}. \label{eq:density_cost_component_exact} \end{equation} Since $\binom{k}{2}=O(k^2)$, the corresponding asymptotic expression is \begin{equation} C_{\mathrm{den},M,\epsilon} = O\!\left( N_{\mathrm{den},M,\epsilon}k^2 \right). \label{eq:density_cost_component_asymptotic} \end{equation}

Finally, $C_{\mathrm{update},M,\epsilon}$ accounts for persistent method-specific state-maintenance operations that are not included in candidate processing. For ACO-based methods, these operations include pheromone initialization, evaporation, and reinforcement. Candidate-score formation, including connectivity-based score computation, probabilistic vertex selection, and executed insertions or removals are counted under $C_{\mathrm{cand},M,\epsilon}$. For methods that do not maintain a persistent search state, $C_{\mathrm{update},M,\epsilon}=0$.

For each graph instance $g$, all GBS-based methods require a one-time graph-to-Gaussian-state mapping. Under a direct dense-matrix implementation, the eigendecomposition and covariance-matrix computation dominate this preprocessing, giving \begin{equation} C_{\mathrm{map},g} = O(n^3). \label{eq:gbs_mapping_cost} \end{equation} This cost is incurred once per graph and reused for all GBS samples. Accordingly, \begin{equation} C_{\mathrm{pre},M,\epsilon} = C_{\mathrm{map},g} + C_{\mathrm{aux},M,\epsilon}, \qquad M\in\mathcal{M}_{\mathrm{GBS}}, \label{eq:gbs_preprocessing_cost} \end{equation} where $C_{\mathrm{aux},\mathrm{GBS+Deg},\epsilon}=O(n+|E|)$ accounts for degree computation, while $C_{\mathrm{aux},M,\epsilon}=0$ for the other GBS-based methods.

For the GBS-based methods, let $N_{r,M,\epsilon}$ denote the number of accepted sample occurrences of cardinality $k+r$ processed before the target gap is attained or the prescribed resource budget is exhausted, where $r\in\{-\Delta,\ldots,\Delta\}$ and $\Delta$ is the maximum accepted cardinality deviation. Repeated sample occurrences are retained. The total number of accepted near-$k$ samples is \begin{equation} N_{\mathrm{near},M,\epsilon} = \sum_{r=-\Delta}^{\Delta} N_{r,M,\epsilon}, \label{eq:near_sample_count} \end{equation} while the number of exact-$k$ samples is \begin{equation} N_{\mathrm{exact},M,\epsilon} = N_{0,M,\epsilon}. \label{eq:exact_sample_count} \end{equation} A sample of cardinality $k+r$ requires $-r$ insertions when $r<0$, $r$ removals when $r>0$, and no modification when $r=0$. Accordingly, define the total numbers of insertion and removal decisions required by a single-pass repair as \begin{align} R^{-}_{M,\epsilon} &= \sum_{r=-\Delta}^{-1} |r|N_{r,M,\epsilon}, \label{eq:insertion_decision_count}\\ R^{+}_{M,\epsilon} &= \sum_{r=1}^{\Delta} rN_{r,M,\epsilon}. \label{eq:removal_decision_count} \end{align} The total number of single-pass repair decisions is therefore \begin{equation} R_{M,\epsilon} = R^{-}_{M,\epsilon} + R^{+}_{M,\epsilon} = \sum_{r=-\Delta}^{\Delta} |r|N_{r,M,\epsilon}. \label{eq:total_repair_decisions} \end{equation} For an insertion decision applied to a current subset $S_i$, the eligible candidate set is $V\setminus S_i$, and hence \begin{equation} m_i=n-|S_i|. \label{eq:insertion_candidate_count} \end{equation} For a removal decision, the eligible candidate set is $S_i$, and hence \begin{equation} m_i=|S_i|. \label{eq:removal_candidate_count} \end{equation} Thus, insertion and removal decisions generally process different numbers of candidate entries. The value of $\kappa_{i,M}$ in \eqref{eq:candidate_processing_cost} depends on the score used by method $M$ and is specified separately for each post-processing strategy.

The classical cost of post-selection consists of the one time graph to Gaussian-state preprocessing, classification of all raw GBS samples according to their cardinality, and complete density evaluations of the accepted exact-$k$ samples. Hence, \begin{equation} \begin{aligned} C_{\mathrm{PS},\epsilon} = O\!\Big( n^3 + Q_{\mathrm{PS},\epsilon} + N_{0,\mathrm{PS},\epsilon}k^2 \Big). \end{aligned} \label{eq:ps_epsilon_cost} \end{equation}

For degree-based post-processing, vertex degrees are computed once at a cost of $O(n+|E|)$. Since candidate selection uses precomputed degree scores, $\kappa_{i,\mathrm{GBS+Deg}}=O(1)$. An insertion decision examines at most $n-k+\Delta$ candidates, whereas a removal decision examines at most $k+\Delta$ candidates. Therefore, \begin{equation} \begin{aligned} C_{\mathrm{cand},\mathrm{GBS+Deg},\epsilon} &= O\!\Big( (n-k+\Delta)R^{-}_{\mathrm{GBS+Deg},\epsilon} \\ &\qquad + (k+\Delta)R^{+}_{\mathrm{GBS+Deg},\epsilon} \Big). \end{aligned} \label{eq:degree_candidate_cost} \end{equation} Since $R_{M,\epsilon}=R^{-}_{M,\epsilon}+R^{+}_{M,\epsilon}$, $k\leq n$, and $\Delta\leq k$, this simplifies to \begin{equation} C_{\mathrm{cand},\mathrm{GBS+Deg},\epsilon} = O\!\left( nR_{\mathrm{GBS+Deg},\epsilon} \right). \label{eq:degree_candidate_cost_simplified} \end{equation} Including sample classification,one time graph to Gaussian-state preprocessing, degree preprocessing, candidate processing, and density evaluations of the accepted near-$k$ samples gives \begin{equation} \begin{aligned} C_{\mathrm{GBS+Deg},\epsilon} = O\!\Big( &Q_{\mathrm{GBS+Deg},\epsilon} + n^3 + |E| + nR_{\mathrm{GBS+Deg},\epsilon} \\ &+ N_{\mathrm{near},\mathrm{GBS+Deg},\epsilon}k^2 \Big). \end{aligned} \label{eq:degree_epsilon_cost} \end{equation}

For edge-based post-processing, each eligible candidate is scored according to its connectivity with the current subset. For an insertion decision, at most $n-k+\Delta$ candidates are examined, and processing each candidate requires at most $O(k)$ adjacency queries. For a removal decision, at most $k+\Delta$ candidates are examined, and processing each candidate requires $O(k+\Delta)$ adjacency queries. Therefore, \begin{equation} \begin{aligned} C_{\mathrm{cand},\mathrm{GBS+Edge},\epsilon} = O\!\Big( &k(n-k+\Delta) R^{-}_{\mathrm{GBS+Edge},\epsilon} \\ & + (k+\Delta)^2 R^{+}_{\mathrm{GBS+Edge},\epsilon} \Big). \end{aligned} \label{eq:edge_candidate_cost} \end{equation} Since $R_{M,\epsilon}=R^{-}_{M,\epsilon}+R^{+}_{M,\epsilon}$, $k\leq n$, and $\Delta\leq k$, this simplifies to \begin{equation} C_{\mathrm{cand},\mathrm{GBS+Edge},\epsilon} = O\!\left( nkR_{\mathrm{GBS+Edge},\epsilon} \right). \label{eq:edge_candidate_cost_simplified} \end{equation} Each accepted near-$k$ sample produces one feasible candidate whose density is evaluated. Consequently, \begin{equation} \begin{aligned} C_{\mathrm{GBS+Edge},\epsilon} = O\!\Big( &Q_{\mathrm{GBS+Edge},\epsilon} + n^3 + nkR_{\mathrm{GBS+Edge},\epsilon} \\ &+ N_{\mathrm{near},\mathrm{GBS+Edge},\epsilon}k^2 \Big). \end{aligned} \label{eq:edge_epsilon_cost} \end{equation}

For ACO-based post-processing, exact-$k$ samples are evaluated directly, while only off-cardinality samples invoke the ACO repair procedure. Each eligible candidate is scored using its connectivity with the current subset, followed by pheromone-guided probability formation and vertex selection. Let $I_{\mathrm{exec}}\leq I$ denote the number of ACO iterations actually executed. Since each of the $a$ ants independently constructs a complete repaired candidate in every executed iteration, the candidate-processing cost is \begin{equation} \begin{aligned} C_{\mathrm{cand},\mathrm{GBS+ACO},\epsilon} = O\!\Big( &I_{\mathrm{exec}}ak(n-k+\Delta) R^{-}_{\mathrm{GBS+ACO},\epsilon} \\ &+ I_{\mathrm{exec}}a(k+\Delta)^2 R^{+}_{\mathrm{GBS+ACO},\epsilon} \Big). \end{aligned} \label{eq:aco_candidate_cost} \end{equation} Since $R_{M,\epsilon}=R^{-}_{M,\epsilon}+R^{+}_{M,\epsilon}$, $k\leq n$, and $\Delta\leq k$, this simplifies to \begin{equation} C_{\mathrm{cand},\mathrm{GBS+ACO},\epsilon} = O\!\left( I_{\mathrm{exec}}ank R_{\mathrm{GBS+ACO},\epsilon} \right). \label{eq:aco_candidate_cost_simplified} \end{equation} Let \begin{equation} N_{\mathrm{off},M,\epsilon} = N_{\mathrm{near},M,\epsilon} - N_{0,M,\epsilon} \label{eq:off_sample_count} \end{equation} denote the number of accepted off-cardinality samples. For each such sample, pheromone initialization requires $O(n)$ operations. Evaporation and pheromone-bound enforcement across $I_{\mathrm{exec}}$ iterations require $O(I_{\mathrm{exec}}n)$ operations. Since only the best candidate is reinforced in each iteration, reinforcement requires $O(I_{\mathrm{exec}}k)$ operations. Therefore, \begin{equation} \begin{aligned} C_{\mathrm{update},\mathrm{GBS+ACO},\epsilon} = O\!\Big( N_{\mathrm{off},\mathrm{GBS+ACO},\epsilon} \big[ &(I_{\mathrm{exec}}+1)n \\ &+ I_{\mathrm{exec}}k \big] \Big). \end{aligned} \label{eq:aco_update_cost} \end{equation} Each exact-$k$ sample is evaluated once. Every off-cardinality sample produces at most $I_{\mathrm{exec}}a$ repaired candidates because every ant constructs one feasible candidate in each executed iteration. Hence, the density-evaluation cost is \begin{equation} \begin{aligned} C_{\mathrm{den},\mathrm{GBS+ACO},\epsilon} = O\!\Big( &N_{0,\mathrm{GBS+ACO},\epsilon}k^2 \\ &+ I_{\mathrm{exec}}a N_{\mathrm{off},\mathrm{GBS+ACO},\epsilon}k^2 \Big). \end{aligned} \label{eq:aco_density_cost} \end{equation} Combining sample classification,one time graph to Gaussian-state preprocessing, candidate processing, density evaluation, and pheromone updates gives \begin{equation} \begin{aligned} C_{\mathrm{GBS+ACO},\epsilon} = O\!\Big( &Q_{\mathrm{GBS+ACO},\epsilon}  +I_{\mathrm{exec}}ank R_{\mathrm{GBS+ACO},\epsilon} \\ &+ N_{\mathrm{off},\mathrm{GBS+ACO},\epsilon} \big[ (I_{\mathrm{exec}}+1)n + I_{\mathrm{exec}}k \big] \\ & + n^3 + I_{\mathrm{exec}}a N_{\mathrm{off},\mathrm{GBS+ACO},\epsilon} k^2 \\ &+ N_{0,\mathrm{GBS+ACO},\epsilon}k^2 \Big). \end{aligned} \label{eq:aco_epsilon_cost} \end{equation}  

Since every off-cardinality sample requires at least one repair decision, $R_{\mathrm{GBS+ACO},\epsilon}\geq N_{\mathrm{off},\mathrm{GBS+ACO},\epsilon}$. Moreover, $n\geq k$ and $I_{\mathrm{exec}},a\geq1$. Consequently, the candidate-processing term asymptotically dominates the pheromone-update and repaired-candidate density-evaluation terms. The total cost therefore simplifies to \begin{equation} \begin{aligned} C_{\mathrm{GBS+ACO},\epsilon} = O\!\Big( &Q_{\mathrm{GBS+ACO},\epsilon} + I_{\mathrm{exec}}ank R_{\mathrm{GBS+ACO},\epsilon} \\ & + n^3 + N_{0,\mathrm{GBS+ACO},\epsilon}k^2 \Big). \end{aligned} \label{eq:aco_epsilon_cost_simplified} \end{equation}

Uniform sampling independently constructs and evaluates $N_{\mathrm{Uni},\epsilon}$ $k$-vertex subsets. Constructing each subset requires $O(k)$ operations, while its complete density evaluation requires $\binom{k}{2}$ pairwise adjacency checks. Therefore, \begin{equation} \begin{aligned} C_{\mathrm{Uni},\epsilon} &= O\!\left( N_{\mathrm{Uni},\epsilon}k + N_{\mathrm{Uni},\epsilon}\binom{k}{2} \right) \\ &= O\!\left( N_{\mathrm{Uni},\epsilon}k^2 \right). \end{aligned} \label{eq:uniform_epsilon_cost} \end{equation}

Degree-based sampling computes vertex degrees once in $O(n+|E|)$ time. When the partial subset contains $j$ vertices, the remaining $n-j$ vertices are processed to form the degree-weighted selection distribution. Thus, the candidate-processing cost per sample is \begin{equation} \begin{aligned} C_{\mathrm{cand},\mathrm{Deg}} &= O(nk). \end{aligned} \label{eq:degree_sampler_candidate_cost} \end{equation} Including the density evaluation of each constructed candidate gives \begin{equation} \begin{aligned} C_{\mathrm{Deg},\epsilon} = O\!\Big( &n+|E| + N_{\mathrm{Deg},\epsilon} \left[nk+k^2\right] \Big). \end{aligned} \label{eq:degree_sampler_epsilon_cost} \end{equation} Since $k\leq n$ implies $k^2\leq nk$, this simplifies to \begin{equation} C_{\mathrm{Deg},\epsilon} = O\!\left( n+|E|+N_{\mathrm{Deg},\epsilon}nk \right). \label{eq:degree_sampler_epsilon_cost_simplified} \end{equation}

Edge-based sampling scores eligible vertices according to their connectivity with the partially constructed subset. When the current subset contains $j$ vertices, the remaining $n-j$ candidates are each compared with $j$ selected vertices. Therefore, the candidate-processing cost per sample is \begin{equation} \begin{aligned} C_{\mathrm{cand},\mathrm{Edge}} &= O(nk^2). \end{aligned} \label{eq:edge_sampler_candidate_cost} \end{equation} Including the complete density evaluation of each constructed candidate gives \begin{equation} C_{\mathrm{Edge},\epsilon} = O\!\left( N_{\mathrm{Edge},\epsilon}nk^2 \right). \label{eq:edge_sampler_epsilon_cost} \end{equation}

Charikar's baseline repeatedly removes a minimum-degree vertex until $k$ vertices remain. Under the direct implementation used here, identifying the minimum-degree vertex requires scanning all active vertices. Therefore, the total candidate-processing cost is \begin{equation} \begin{aligned} C_{\mathrm{cand},\mathrm{Char}} &= O\!\left( \sum_{s=k+1}^{n}s \right) \\ &= O(n^2). \end{aligned} \label{eq:charikar_candidate_cost} \end{equation} Removing vertices and maintaining the degrees of their active neighbors contributes $O(|E|)$ operations over the complete peeling process, while evaluating the density of the final $k$-vertex subset contributes $O(k^2)$. Hence, \begin{equation} \begin{aligned} C_{\mathrm{Char}} = O\!\Big( &n^2 + |E| + k^2 \Big). \end{aligned} \label{eq:charikar_epsilon_cost} \end{equation}

In classical ACO, each of the $a$ ants constructs one $k$-vertex candidate during each of the $I$ iterations. When the partial subset contains $j$ vertices, the remaining $n-j$ candidates are each scored against the $j$ selected vertices. Therefore, the candidate-processing cost is \begin{equation} \begin{aligned} C_{\mathrm{cand},\mathrm{ACO}}  &= O(Iank^2). \end{aligned} \label{eq:classical_aco_candidate_cost} \end{equation} Each ant produces one complete candidate per iteration, giving the density-evaluation cost \begin{equation} C_{\mathrm{den},\mathrm{ACO}} = O(Iak^2). \label{eq:classical_aco_density_cost} \end{equation} Pheromone initialization requires $O(n)$ operations, while evaporation and pheromone-bound enforcement require $O(In)$ operations. If only the best candidate is reinforced in each iteration, reinforcement requires $O(Ik)$ operations. Hence, \begin{equation} \begin{aligned} C_{\mathrm{update},\mathrm{ACO}} = O\!\Big( &n + In + Ik \Big). \end{aligned} \label{eq:classical_aco_update_cost} \end{equation} Combining candidate processing, density evaluation, and pheromone updates gives \begin{equation} \begin{aligned} C_{\mathrm{ACO}} = O\!\Big( &Iank^2 + Iak^2 + In + Ik + n \Big). \end{aligned} \label{eq:classical_aco_cost} \end{equation} Since $n\geq k$ and $I,a\geq1$, the candidate-processing term dominates, and the total cost simplifies to \begin{equation} C_{\mathrm{ACO}} = O(Iank^2). \label{eq:classical_aco_epsilon_cost} \end{equation}

Brute-force search enumerates every possible $k$-vertex subset. Hence, the number of candidate subsets is \begin{equation} N_{\mathrm{BF}} = \binom{n}{k}. \label{eq:brute_force_candidate_count} \end{equation} Enumerating each subset requires $O(k)$ operations, while evaluating its density requires $\binom{k}{2}$ pairwise adjacency checks. Therefore, \begin{equation} \begin{aligned} C_{\mathrm{BF}} &= O\!\left( \binom{n}{k} \left[ k+\binom{k}{2} \right] \right) \\ &= O\!\left( \binom{n}{k}k^2 \right). \end{aligned} \label{eq:brute_force_cost} \end{equation} 

The preceding expressions follow from the dominant operations performed by each method. Since a method may not attain the prescribed target optimality gap $\epsilon$ within a practical resource budget, each method is evaluated under fixed quantum and classical resource limits. For graph instance $g$, the resource vector \begin{equation} \mathcal{R}_{M,\epsilon}^{(g)} = \left( Q_{M,\epsilon}^{(g)}, C_{M,\epsilon}^{(g)} \right) \end{equation} is recorded when the target condition $\epsilon_{M,g}\leq\epsilon$ is first satisfied. The censored quantum cost is defined as \begin{equation} Q_{M,\epsilon}^{(g),\mathrm{cens}} = \begin{cases} Q_{M,\epsilon}^{(g)}, & \text{if the target is reached}, \\[2mm] Q_{M,\mathrm{budget}}^{(g)}, & \text{otherwise}, \end{cases} \label{eq:censored_quantum_cost} \end{equation} and the censored classical cost is \begin{equation} C_{M,\epsilon}^{(g),\mathrm{cens}} = \begin{cases} C_{M,\epsilon}^{(g)}, & \text{if the target is reached}, \\[2mm] C_{M,\mathrm{budget}}^{(g)}, & \text{otherwise}. \end{cases} \label{eq:censored_classical_cost} \end{equation} If the target is not reached, the resources consumed up to the prescribed budget are reported as censored costs according to Eqs.~\eqref{eq:censored_quantum_cost} and \eqref{eq:censored_classical_cost}. The aggregated quantum and classical resource costs over $G$ graph instances are \begin{equation} Q_{M,\epsilon}^{\mathrm{total}} = \sum_{g=1}^{G} Q_{M,\epsilon}^{(g),\mathrm{cens}}, \label{eq:total_quantum_cost} \end{equation} and \begin{equation} C_{M,\epsilon}^{\mathrm{total}} = \sum_{g=1}^{G} C_{M,\epsilon}^{(g),\mathrm{cens}}. \label{eq:total_classical_cost} \end{equation}

The aggregated quantum and classical resource costs for the synthetic community-structured graphs are reported in Table~\ref{tab:unified_quality_cost}. The above resource model complements the solution-quality metrics used throughout this work and provides a common basis for comparing hybrid and classical approaches. In the following section, we present numerical results assessing convergence, solution quality, robustness, and resource requirements. 

\section{Simulations and Results}\label{sec:simulations_results}
To validate the effectiveness of the proposed enhancement strategies over GBS with post-selection for the D$k$SP, numerical simulations are performed and their performance with classical methods, 
that are discussed in Section~\ref{sec:classical methods}, are compared.

\subsection{Convergence and Performance on Erd\H{o}s--R\'enyi Graphs} \label{sec:er_results}
Experiments are conducted on randomly generated Erd\H{o}s--R\'enyi graphs $G(n,p)$~\cite{erdos_random_graph}, where $n$ is the number of vertices and $p$ is the probability of an edge between two vertices. All simulations are implemented on Python-based \emph{Strawberry Fields} platform~\cite{straw}, an open-source framework for simulating photonic quantum circuits. For each graph instance, the adjacency matrix $A$ was encoded into the GBS device after rescaling as $\tilde{A}=cA$, where $c=0.9/\lambda_{\max}$ and $\lambda_{\max}$ is the largest eigenvalue of $A$. The sampler was initialized with mean photon number $n_{\mathrm{mean}}=10$. Under threshold detection, each output mode gives a binary outcome, where $s_i=0$ denotes no photon detection and $s_i=1$ denotes one or more detected photons. The clicked modes define the sampled vertex subset $\tilde{S}\subseteq V$, which was then used for post-selection and for the proposed post-processing methods.

\begin{figure}[ht]
\centering
\includegraphics[width=1\linewidth]{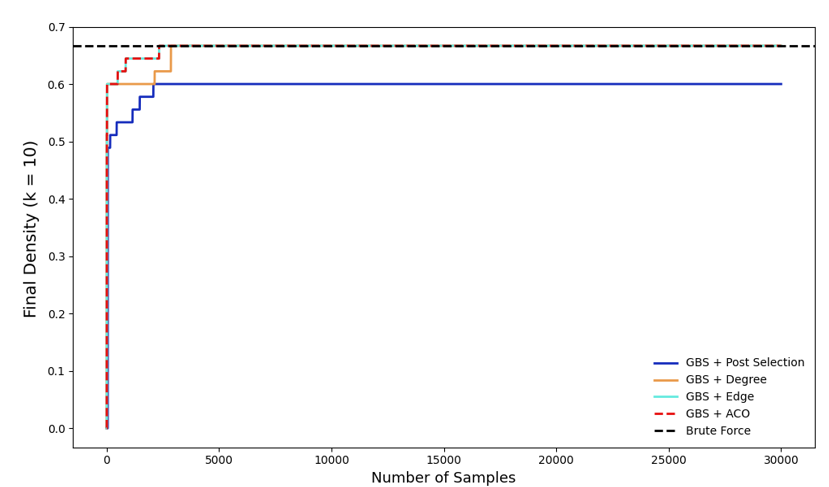}
\caption{Convergence of the final density found for $k=10$ subgraphs on a random graph with edge probability $p=0.3$.}
\label{fig:convergence_plot}
\end{figure}

We first analyze the convergence of the GBS-based samplers with post-selection and post-processing for an Erd\H{o}s--R\'enyi graph with $n=30$, edge probability $p=0.3$, while fixing the subgraph size $k=10$ to check the \textit{sampler efficiency} of the various samplers. The final density obtained for $10$-vertex subgraphs is plotted as a function of the number of samples in Fig.~\ref{fig:convergence_plot}. For each trial sample of GBS, different post-processing is employed to enhance the solution and the resulting density is recorded. For the GBS methods with post-processing, we considered near-$k$ samples with the number of clicks satisfying $|S| \in [k-2,k+2]$, i.e., subgraphs with slightly fewer or more than $k$ vertices. Only these samples are considered as valid and subsequently resized to obtain valid $k=10$ subgraphs using the post-processing strategies. For the ACO-based post-processing, the parameters are set as follows: number of ants $m=2$, a maximum $20$ iterations, $\tau_0 = 0.1$, $\rho = 0.9$, and $\delta = 1$.  The \emph{brute-force} solution is also added for reference but with the high computational cost of checking for all the possible $10$-vertex subgraphs (i.e., $\binom{30}{10}$).
GBS with edge-based and ACO-based post-processing converges fastest, reaching the optimal solution within $\sim2000$ samples. For early samples, GBS with hard post-selection produces no accepted $k$-vertex sample in the early sampling regime, so the best valid density remains zero until the first feasible sample is observed. Furthermore, it fails to reach the optimal results even after 
$50000$ samples, whereas GBS with any of the three post-processing reached the optimal value. Similar trends were also observed for other choices of $p$, $n$, and $k$, where edge-based or ACO-based post-processing consistently reached the optimal or near-optimal density, while post-selection required substantially more samples. These results indicate that post-processing plays a critical role in extracting dense structures from direct GBS samples indicating a need of GBS as a sub-routine sampler.

\begin{figure}[ht]
\centering
\includegraphics[width=1\linewidth]{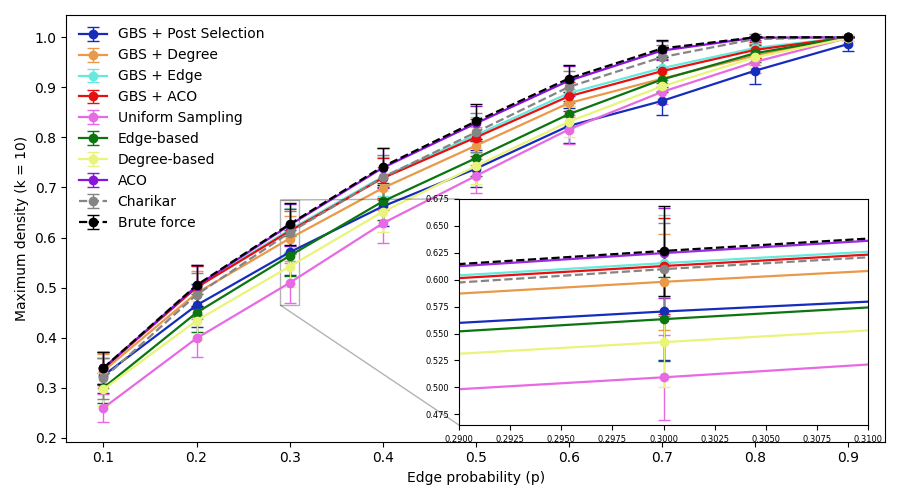}
\caption{Comparison of sampling and classical methods for DkSP with $k=10$  in Erd\H{o}s--R\'enyi random graphs with $n=30$. Results are averaged over 100 random graph instances for each edge probability $p$.}\label{fig:avg_plot_random_graphs}
\end{figure}

To assess robustness across graph regimes, for each $p \in \{0.1,0.2,\ldots,0.9\}$ we sample $100$ independent graphs $G \sim \mathcal{G}(30,p)$. The total number of samples is fixed to $2000$ for all the samplers. The average maximum density over $100$ graphs of $10$-vertex subgraphs is plotted against $p$ in  Fig.~\ref{fig:avg_plot_random_graphs}. Methods include GBS-based approaches: degree-based, edge-based, and ACO-based (with $m=1$ and $5$ maximum iterations), classical baselines (ACO with ants $m=20$ and $100$ iterations, brute force and Charikar), and classical sampling strategies (uniform, degree-based, and edge-based) with $2000$ total samples. For the GBS methods with post-processing, we considered near-$k$ samples with the number of clicks satisfying $|S| \in [k-2,k+2]$, i.e., subgraphs with slightly fewer or more than $k$ vertices. Only these samples are considered as valid and subsequently resized to obtain valid $k=10$ subgraphs using the post-processing strategies. ACO achieves near-optimal performance due to its effectiveness as a classical metaheuristic. In the sparse regime ($p \leq 0.4$), GBS combined with edge-based and ACO-based post-processing outperforms Charikar. GBS with edge and ACO-based post-processing consistently outperformed the other samplers in the sparse regime. 
The average of maximum density increases as $p$ increases, reflecting the increased availability of dense subgraphs in denser graphs (high $p$). In the highly dense regime ($p\geq0.7$), \emph{uniform sampling} outperformed post-selection GBS because of the higher probability of obtaining a dense subgraph with increased $p$. However, edge-based and aco-based post-processing, performed well in both sparse and dense regimes.
This shows that allowing a small tolerance around the target subgraph size and then refining the sampled subgraph can significantly improve the density of solution from GBS for D$k$SP. 

\subsection{Robustness to Photon Loss} \label{sec:loss_results}
\begin{figure}[ht] \centering \includegraphics[width=\columnwidth]{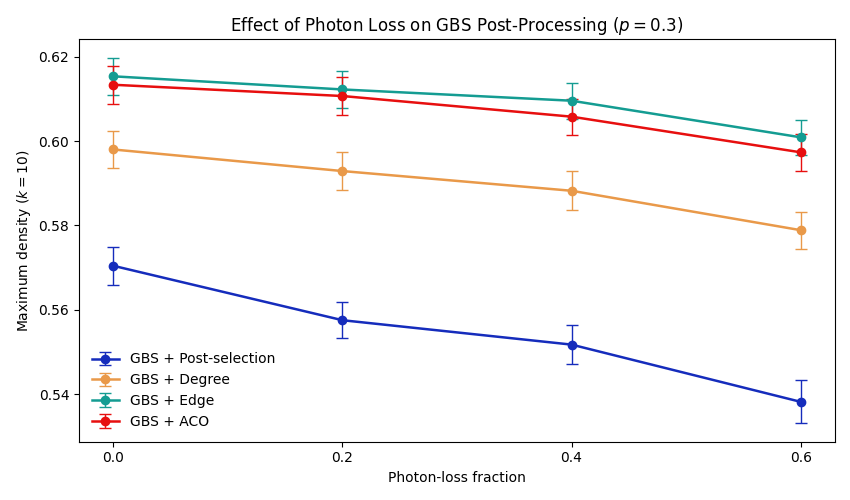} \caption{Effect of photon loss on GBS-based D$k$SP for $k=10$ on Erd\H{o}s--R\'enyi graphs with $p=0.3$. The curves show the average maximum density obtained using hard post-selection and the proposed post-processing methods.} \label{fig:loss_density_p03} \end{figure}

\begin{figure}[ht] \centering \includegraphics[width=\columnwidth]{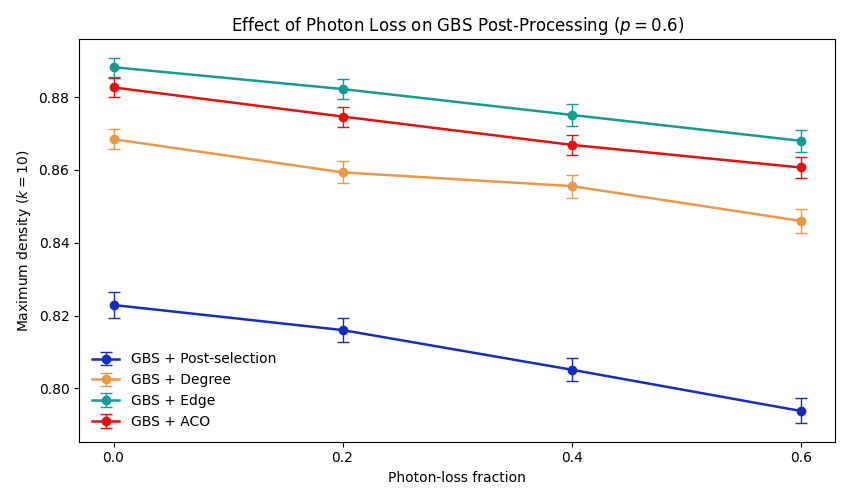} \caption{Effect of photon loss on GBS-based D$k$SP for $k=10$ on Erd\H{o}s--R\'enyi graphs with $p=0.6$.} \label{fig:loss_density_p06} \end{figure}

To evaluate the robustness of the proposed methods under a realistic photonic implementation, we introduce photon-loss fractions $\ell\in\{0,0.2,0.4,0.6\}$ in the GBS sampler, where $\ell=0$ denotes ideal loss-free sampling. The loss is applied during GBS sample generation before threshold detection. For each combination of $p$ and $\ell$, we generate $2000$ raw GBS samples for each of $100$ independently generated Erd\H{o}s--R\'enyi graphs, while fixing $n=30$ and $k=10$. The same near-$k$ acceptance interval, $|S|\in[k-2,k+2]$, and post-processing parameters are used for all loss levels. Fig.~\ref{fig:loss_density_p03} and~\ref{fig:loss_density_p06} show the average maximum density as a function of photon loss for $p=0.3$ and $p=0.6$, representing sparse and moderately dense graph regimes, respectively. Increasing photon loss generally reduces the probability of observing samples close to the target cardinality. Consequently, the performance of hard post-selection becomes less reliable under higher loss. In contrast, the proposed post-processing methods continue to exploit near-$k$ samples and maintain higher solution densities. Edge-based and ACO-based post-processing provide the strongest performance across the tested loss levels.   

\begin{figure}[ht] \centering \includegraphics[width=\columnwidth]{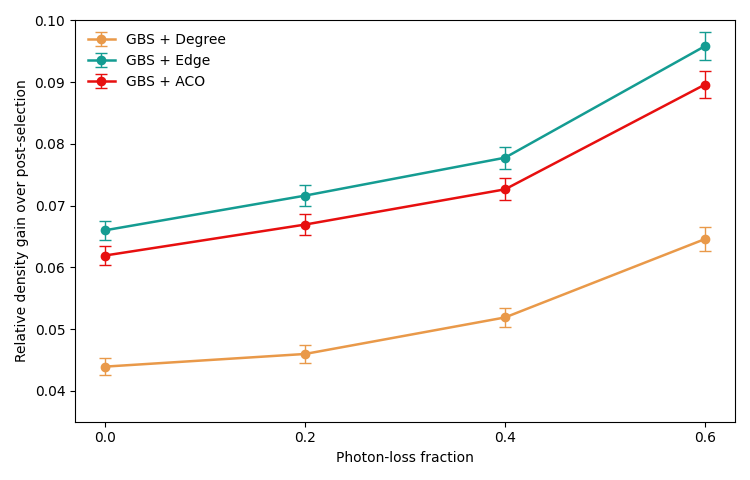} \caption{Mean relative density gain of the proposed post-processing methods over hard GBS post-selection under different photon-loss fractions for $k=10$.} \label{fig:loss_gain} \end{figure}

To quantify the benefit of post-processing more directly, we define the relative density gain for method $M$ as \begin{equation} \Delta D_M = \frac{D_M-D_{\mathrm{PS}}}{D_{\mathrm{PS}}}, \end{equation} where $D_M$ is the maximum density obtained using method $M$, and $D_{\mathrm{PS}}$ is the maximum density obtained using hard post-selection on the same graph under the same photon-loss level and sample budget. Fig.~\ref{fig:loss_gain} reports the mean relative density gain, averaged over $p\in\{0.1,0.2,\ldots,0.9\}$. A positive value of $\Delta D_M$ indicates that post-processing improves upon hard post-selection. All three methods produce a positive average percentage gain across the tested loss levels, with edge-based and ACO-based post-processing generally providing the largest improvements. The gain remains positive as photon loss increases, showing that near-$k$ samples retain useful structural information even when exact-$k$ samples become less frequent. These results indicate that classical refinement improves the robustness of the GBS-based framework to photon loss. Relative gains are computed only for instances in which hard post-selection produces a non-zero density solution, since the relative gain is undefined when $D_{\mathrm{PS}}=0$. 

\subsection{Performance on Community-Structured Graphs} \label{sec:community_results}
While Erd\H{o}s--R\'enyi graphs provide an unstructured baseline, many real-world networks exhibit community or modular structure~\cite{modular_graphs}, where vertices form densely connected groups with sparser connections between groups. To evaluate the proposed methods on structured graphs inspired by real-world community networks, we synthesize community-structured graphs.

\begin{figure}[ht]
\centering
\includegraphics[width=1\linewidth]{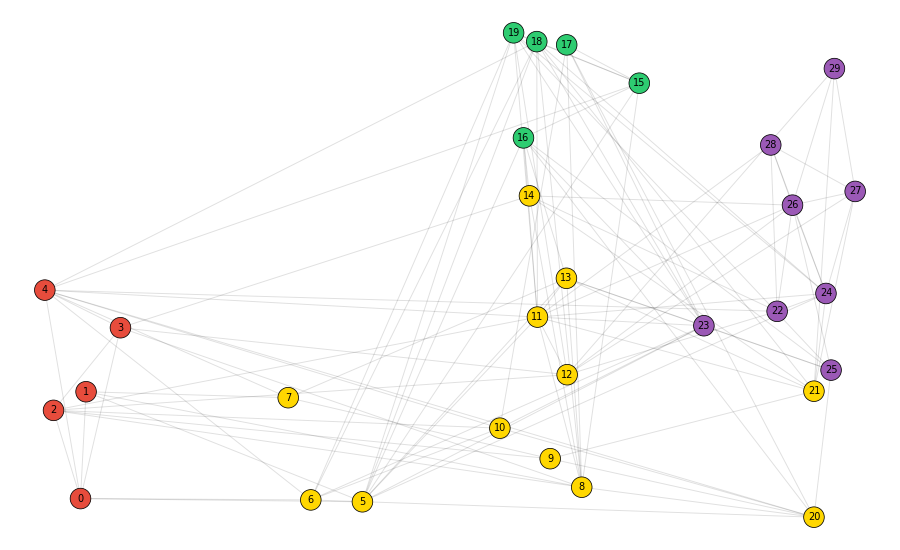}
\caption{Synthetic graph with overlapping dense regions for $n=30$. Colors indicate cluster membership, and overlapping nodes are highlighted in gold.}
\label{fig:multiple_dense_graph}
\end{figure}

Each graph consists of $n=30$ vertices with background edge probability $p_{\mathrm{out}}=0.1$ and contains four overlapping moderately dense clusters with edge probability $p_{\mathrm{in}}=0.5$: $C_1=\{0,\dots,9\}$, $C_2=\{5,\dots,14\}$, $C_3=\{8,\dots,21\}$, and $C_4=\{20,\dots,29\}$. This construction produces graphs with sparse inter-cluster connectivity and denser intra-cluster regions, as illustrated in Fig.~\ref{fig:multiple_dense_graph}.

This construction is inspired by planted dense-subgraph and community-structured models, but differs from the standard single planted subgraph setting, since it contains multiple overlapping dense regions that create competing high-density solutions~\cite{modular_graphs}.

\begin{figure}[ht]
\centering
\includegraphics[width=1\linewidth]{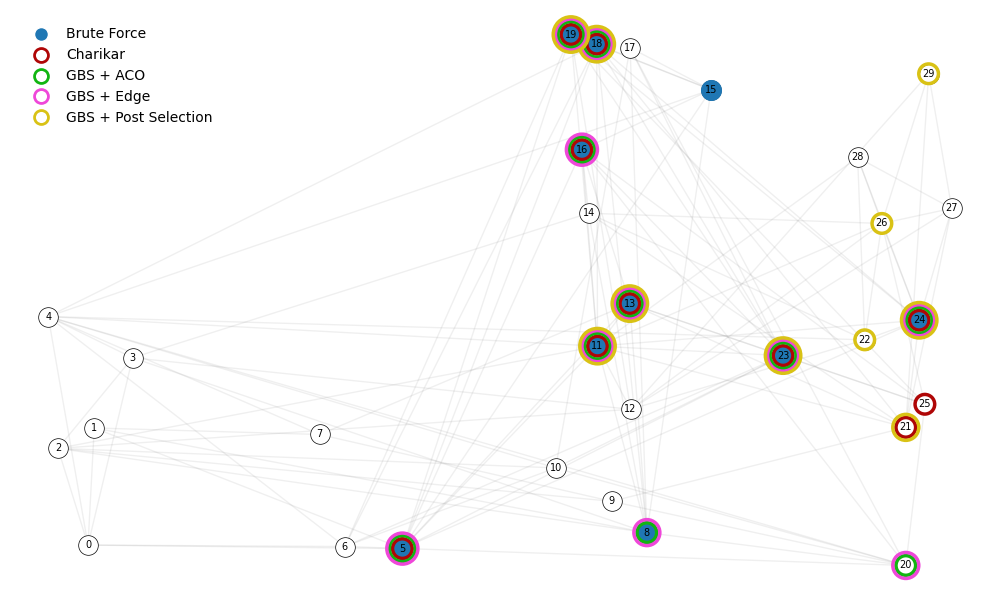}
\caption{Dense $k=10$ subgraphs identified by different methods on the same multiple dense-region graph. The selected nodes are overlaid on a common layout for comparison.}
\label{fig:subgraphs_multiple_dense}
\end{figure}
Fig.~\ref{fig:subgraphs_multiple_dense} visualizes the solutions obtained by different methods on a representative instance of the multiple dense-region graph. The GBS solution with post-selection differs from the brute-force optimum by four vertices, whereas Charikar's solution differs by two vertices. In contrast, GBS with edge-based and ACO-based post-processing select the same subgraph and differ from the brute-force optimum by only one vertex. This demonstrates that post-processing significantly improves the quality of GBS samples and provides a closer approximation to the optimal dense subgraph than post-selection alone, even in structured graphs with multiple competing dense regions.

\begin{figure}[ht] \centering \includegraphics[width=1\linewidth]{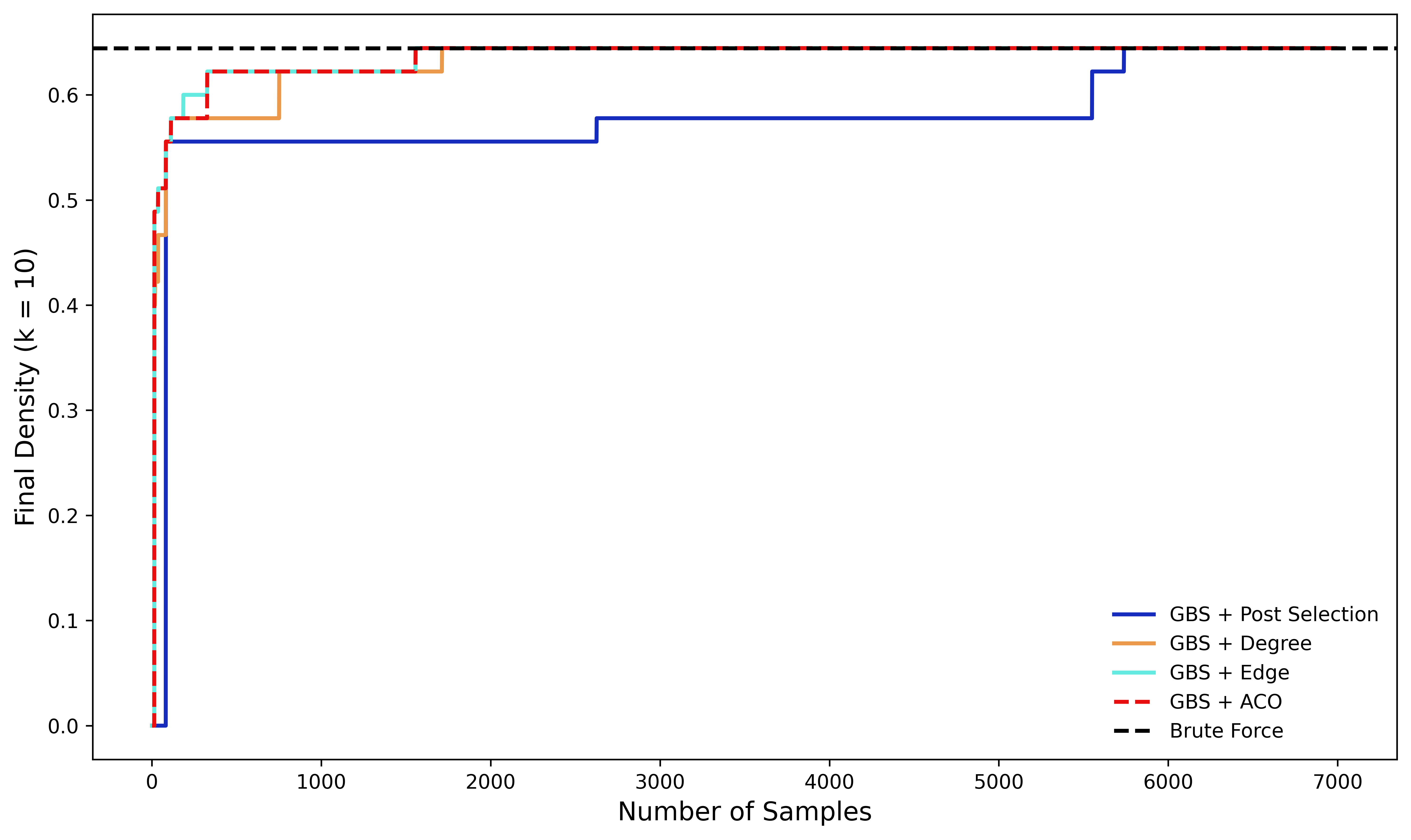} \caption{Convergence of the final density found for $k=10$ subgraphs on a multiple dense-region graph.}\label{fig:convergence_plot_multiple_dense_region} \end{figure}

We next evaluate convergence on community-structured graphs, analogous to the Erd\H{o}s--R\'enyi case in Fig.~\ref{fig:convergence_plot}. As shown in Fig.~\ref{fig:convergence_plot_multiple_dense_region}, post-processing significantly accelerates convergence. In particular, edge-based and ACO-based methods reach the optimum within $\sim1500$ samples, whereas hard post-selection requires $\sim6000$ samples. Thus, GBS with post-processing brings $4$-times more sampling efficiency than GBS with post-selection. This demonstrates that the proposed post-processing strategies are robust across both unstructured and structured graphs, and significantly improve the effectiveness of GBS as a sampling-based approach for the D$k$SP.
 
We also evaluate the average performance over $100$ independently generated community-structured graphs, as described above, with $n=30$. Table~\ref{tab:multi_dense} reports the maximum density achieved for different subgraph with sizes $k$ from $6$ to $11$. The GBS-based methods with post-processing consistently outperform post-selection and classical sampling methods. Although GBS with post-selection performs better than uniform sampling, the improvement is relatively modest compared with degree-based and edge-based classical sampling. In contrast, GBS with edge-based and ACO-based post-processing achieves the best performance among the sampling-based methods and closely approaches the brute-force optimum across all values of $k$. This improvement over post-selection shows that even simple post-processing strategies can significantly enhance the sampling efficiency and solution quality of GBS.
\begin{table}[ht]
\centering
\caption{Benchmarking different methods for D$k$SP on community-structured graphs.}
\label{tab:multi_dense}
\begin{tabular}{lcccccc}
\hline
Method & \multicolumn{6}{c}{Maximum density for subgraph size $k$} \\
\cline{2-7}
 & 6 & 7 & 8 & 9 & 10 
 \\
\hline
GBS + Post-selection & 0.830 & 0.759 & 0.698 & 0.651 & 0.611 
\\
GBS + Degree & 0.869 & 0.793 & 0.743 & 0.685 & 0.642 
\\
GBS + Edge & \textbf{0.891} & 0.818 & 0.758 & 0.703 & 0.663 
\\
GBS + ACO & 0.890 & 0.814 & 0.756 & 0.701 & 0.659 
\\

\hline
Uniform Sampling & 0.733 & 0.662 & 0.604 & 0.566 & 0.533 
\\
Degree-based sampling & 0.787 & 0.714 & 0.662 & 0.616 & 0.580 
\\
Edge-based sampling & 0.847 & 0.771 & 0.704 & 0.665 & 0.618 
\\
\hline
Charikar & 0.693 & 0.790 & 0.553 & 0.693 & 0.652 
\\
ACO & 0.880 & 0.812 & 0.757 &  0.706 & 0.660 
\\
Brute Force & \textbf{0.891} & \textbf{0.820} & \textbf{0.759} & \textbf{0.708} & \textbf{0.665} 
\\
\hline
\end{tabular}
\end{table} 

\subsection{Resource and Cost Comparison} \label{sec:cost_results}
The total quantum and classical costs are aggregated over all graph instances using Eqs.~(\ref{eq:total_quantum_cost}) and (\ref{eq:total_classical_cost}). Table~\ref{tab:unified_quality_cost} reports results obtained with a maximum budget of $5{,}000$ samples per graph instance for the classical sampling methods and $2{,}000$ raw GBS samples per graph instance for all GBS-based methods. GBS with ACO-based post-processing uses $1$ ant and $5$ iterations, whereas the classical ACO baseline uses $20$ ants and $100$ iterations.

For GBS-based methods, the reported classical runtime includes only classical processing and excludes GBS sample-generation time. Since GBS is numerically simulated in this work, including its simulation time would measure classical emulation rather than physical quantum sampling. Quantum sampling and classical computation are therefore reported separately.

Among the hybrid methods, GBS+Edge requires the lowest quantum cost, consuming $95{,}280$ raw GBS samples while attaining a mean gap of $0.67\%$. GBS+ACO requires $131{,}072$ GBS samples and approximately $39.96$ million classical operations, attaining a mean gap of $1.88\%$. Despite the additional post-processing, its classical workload remains below that of classical ACO, which requires approximately $46.91$ million operations and attains a mean gap of $1.45\%$. The corresponding measured classical runtimes are $8.08$~s and $11.77$~s, respectively. GBS+Edge incurs substantially less classical computation, approximately $8.12$ million operations and $1.08$~s of classical runtime, while attaining the smallest mean gap among the non-exhaustive methods shown in Table~\ref{tab:unified_quality_cost}.

Overall, the results show a trade-off between quantum-sample efficiency, classical processing cost, and solution quality. Edge-based post-processing provides the best quality per GBS sample, while ACO-based refinement incurs greater classical overhead than the greedy methods but, under the evaluated configuration, remains less costly than classical ACO. Classical runtime follows the same general trend as the operation counts and is reported only as an implementation-dependent supporting metric.
\begin{table*}[t]
\centering
\scriptsize
\caption{Quality-normalized resource comparison for $k=10$ over $100$ community-structured graph instances at target gap $\epsilon=0.005$. For GBS-based methods, the reported runtime includes only classical processing and excludes GBS sample-generation time.}
\label{tab:unified_quality_cost}
\resizebox{\textwidth}{!}{%
\begin{tabular}{lrrrcc} \hline Method & Quantum cost & Classical cost & Classical Runtime (s) & Mean gap (\%) & Asymptotic cost \\ \hline Uniform sampling & -- & 27,500,000 & 7.6250 & 16.06 & Eq.~\eqref{eq:uniform_epsilon_cost} \\ Degree-based sampling & -- & 550,276,400 & 37.7198 & 10.33 & Eq.~\eqref{eq:degree_sampler_epsilon_cost} \\ Edge-based sampling & -- & 926,796,400 & 190.8824 & 5.35 & Eq.~\eqref{eq:edge_sampler_epsilon_cost} \\ \hline GBS + Post-selection & 192,416 & 3,227,261 & 0.1926 & 8.20 & Eq.~\eqref{eq:ps_epsilon_cost} \\ GBS + Degree & 162,330 & 5,074,202 & 0.7064 & 3.40 & Eq.~\eqref{eq:degree_epsilon_cost} \\ GBS + Edge & 95,280 & 8,119,256 & 1.0835 & 0.67 & Eq.~\eqref{eq:edge_epsilon_cost} \\ GBS + ACO & 131,072 & 39,959,520 & 8.0789 & 1.88 & Eq.~\eqref{eq:aco_epsilon_cost} \\ \hline Charikar & -- & 176,064 & 0.0676 & 1.97 & Eq.~\eqref{eq:charikar_epsilon_cost} \\ ACO & -- & 46,909,310 & 11.7681 & 1.45 & Eq.~\eqref{eq:classical_aco_epsilon_cost} \\ Brute-force & -- & 165,247,582,500 & 30681.7570 & 0.00 & Eq.~\eqref{eq:brute_force_cost}\\ \hline \end{tabular}%
}
\end{table*}

\subsection{Performance on Real-World Networks} \label{sec:realworld_results}
To assess the performance of the proposed methods on real-world networks, we consider the Zachary Karate Club social network with $34$ vertices and $78$ friendship edges~\cite{zachary}, the Dolphin social network with $62$ vertices and $159$ association edges~\cite{dolphin_data}, and the Political Books network with $105$ vertices and $441$ co-purchasing edges~\cite{books_data}. Identifying dense sub-graphs in these networks is relevant to the discovery of strongly interacting groups, highly connected communities, and strongly associated entities in social, information, and recommendation networks. On the Karate Club network, using $2000$ GBS samples and the near-cardinality condition $\lvert |\tilde{S}|-k\rvert\leq2$, the proposed post-processing methods closely match the brute-force optimum for $k=7,\ldots,15$ and recover the optimum for all tested values except $k=10$, where refinement improves the density from $0.489$ to $0.533$, compared with the optimum of $0.556$. For this dataset, the classical ACO baseline uses $20$ ants and $100$ iterations, whereas GBS+ACO uses only $1$ ant and $5$ iterations. On the Dolphin network, using $20000$ GBS samples and $\lvert |\tilde{S}|-k\rvert\leq3$, GBS+Edge and GBS+ACO achieve the highest or joint-highest densities among the GBS-based methods and outperform the Charikar baseline for $k=17,\ldots,30$. Here, classical ACO uses $40$ ants and $500$ iterations, compared with only $2$ ants and $25$ iterations for GBS+ACO. On the Political Books network, using $30000$ GBS samples and $\lvert |\tilde{S}|-k\rvert\leq5$, the proposed methods consistently improve upon hard post-selection and continue to produce valid high-density solutions for $k\geq31$, where no exact-$k$ sample is observed within the sampling budget. For this dataset, classical ACO uses $60$ ants and $500$ iterations, whereas GBS+ACO uses $3$ ants and $25$ iterations. Although classical ACO often attains the highest densities on the Dolphin and Political Books networks, the proposed refinement methods substantially improve upon hard post-selection while requiring significantly smaller optimization budgets. Overall, the results demonstrate that off-cardinality GBS samples retain useful graph-structural information and that lightweight near-$k$ post-processing improves effective sampling coverage and D$k$SP solution quality across diverse real-world networks.

\begin{table}[ht] \centering \small \caption{Final density obtained on the Zachary Karate Club network for different subgraph sizes $k$.} \label{tab:karate} \resizebox{\columnwidth}{!}{ \begin{tabular}{cccccccc} \hline $k$ & GBS+PS & GBS+Degree & GBS+Edge & GBS+ACO & Charikar & ACO & Brute Force\\ \hline 7 & \textbf{0.762} & \textbf{0.762} & \textbf{0.762} & \textbf{0.762} & 0.619 & \textbf{0.762} & \textbf{0.762}\\ 8 & \textbf{0.643} & \textbf{0.643} & \textbf{0.643} & \textbf{0.643} & 0.607 & \textbf{0.643} & \textbf{0.643}\\ 9 & \textbf{0.583} & \textbf{0.583} & \textbf{0.583} & \textbf{0.583} & \textbf{0.583} & \textbf{0.583} & \textbf{0.583}\\ 10 & 0.489 & 0.533 & 0.533 & 0.533 & \textbf{0.556} & \textbf{0.556} & \textbf{0.556}\\ 11 & 0.491 & \textbf{0.509} & \textbf{0.509} & \textbf{0.509} & 0.491 & \textbf{0.509} & \textbf{0.509}\\ 12 & 0.439 & \textbf{0.470} & \textbf{0.470} & \textbf{0.470} & 0.455 & \textbf{0.470} & \textbf{0.470}\\ 13 & 0.410 & 0.423 & \textbf{0.436} & \textbf{0.436} & 0.423 & \textbf{0.436} & \textbf{0.436}\\ 14 & 0.385 & \textbf{0.396} & \textbf{0.396} & \textbf{0.396} & \textbf{0.396} & \textbf{0.396} & \textbf{0.396}\\ 15 & 0.362 & 0.362 & \textbf{0.371} & \textbf{0.371} & 0.362 & \textbf{0.371} & \textbf{0.371}\\ \hline \end{tabular} } \end{table}

\begin{table}[ht] \centering \caption{Final density obtained on the Dolphin social network.} \label{tab:dolphin_results} \resizebox{\columnwidth}{!}{ \begin{tabular}{c|ccccccc} \hline $k$ & GBS+PS & GBS+Degree & GBS+Edge & GBS+ACO & Charikar & ACO \\ \hline 15 & 0.390 & 0.410 & \textbf{0.419} & \textbf{0.419} & \textbf{0.419} & \textbf{0.419} \\ 16 & 0.367 & 0.392 & \textbf{0.400} & \textbf{0.400} & \textbf{0.400} & \textbf{0.400} \\ 17 & 0.375 & 0.375 & \textbf{0.382} & \textbf{0.382} & 0.353 & \textbf{0.382} \\ 18 & 0.320 & \textbf{0.359} & \textbf{0.359} & \textbf{0.359} & 0.320 & \textbf{0.359} \\ 19 & 0.304 & 0.327 & \textbf{0.339} & \textbf{0.339} & 0.298 & \textbf{0.339} \\ 20 & 0.311 & 0.311 & \textbf{0.326} & \textbf{0.326} & 0.284 & \textbf{0.326} \\ 21 & 0.271 & 0.300 & 0.305 & 0.305 & 0.271 & \textbf{0.309} \\ 22 & 0.281 & 0.281 & \textbf{0.294} & \textbf{0.294} & 0.264 & \textbf{0.294} \\ 23 & 0.249 & 0.269 & \textbf{0.281} & \textbf{0.281} & 0.257 & 0.280 \\ 24 & 0.232 & 0.257 & \textbf{0.268} & \textbf{0.268} & 0.250 & \textbf{0.268} \\ 25 & 0.240 & 0.247 & \textbf{0.257} & \textbf{0.257} & 0.240 & 0.256 \\ 26 & 0.206 & 0.237 & 0.237 & 0.237 & 0.231 & \textbf{0.246} \\ 27 & 0.225 & 0.225 & 0.231 & 0.231 & 0.225 & \textbf{0.236} \\ 28 & 0.183 & 0.209 & 0.225 & 0.225 & 0.220 & \textbf{0.227} \\ 29 & 0.170 & 0.197 & 0.217 & 0.217 & 0.212 & \textbf{0.219} \\ 30 & 0.177 & 0.189 & 0.209 & 0.209 & 0.205 & \textbf{0.211} \\ \hline \end{tabular} } \end{table}

\begin{table}[ht] \centering \caption{Final density obtained on the Political Books network.} \label{tab:books_results} \resizebox{\columnwidth}{!}{ \begin{tabular}{c|ccccccc} \hline $k$ & GBS+PS & GBS+Degree & GBS+Edge & GBS+ACO & Charikar & ACO \\ \hline 15 & 0.610 & 0.619 & \textbf{0.638} & \textbf{0.638} & \textbf{0.638} & 0.610 \\ 16 & 0.575 & \textbf{0.608} & \textbf{0.608} & \textbf{0.608} & \textbf{0.608} & \textbf{0.608} \\ 17 & 0.529 & 0.574 & \textbf{0.581} & \textbf{0.581} & \textbf{0.581} & \textbf{0.581} \\ 18 & 0.503 & 0.542 & \textbf{0.549} & \textbf{0.549} & 0.516 & \textbf{0.549} \\ 19 & 0.503 & 0.503 & \textbf{0.520} & \textbf{0.520} & 0.468 & \textbf{0.520} \\ 20 & 0.442 & 0.463 & \textbf{0.495} & 0.489 & 0.432 & \textbf{0.495} \\ 21 & 0.433 & 0.448 & \textbf{0.471} & \textbf{0.471} & 0.405 & 0.448 \\ 22 & 0.403 & 0.424 & \textbf{0.450} & \textbf{0.450} & 0.381 & \textbf{0.450} \\ 23 & 0.387 & 0.411 & \textbf{0.431} & 0.427 & 0.364 & \textbf{0.431} \\ 24 & 0.380 & 0.395 & 0.406 & 0.406 & 0.348 & \textbf{0.413} \\ 25 & 0.343 & 0.370 & 0.387 & 0.387 & 0.337 & \textbf{0.393} \\ 26 & 0.332 & 0.357 & \textbf{0.369} & \textbf{0.369} & 0.329 & 0.363 \\ 27 & 0.296 & 0.342 & \textbf{0.353} & \textbf{0.353} & 0.322 & 0.350 \\ 28 & 0.325 & 0.325 & 0.341 & 0.341 & 0.315 & \textbf{0.349} \\ 29 & 0.276 & 0.310 & 0.328 & 0.328 & 0.308 & \textbf{0.335} \\ 30 & 0.297 & 0.297 & 0.313 & 0.313 & 0.299 & \textbf{0.322} \\ 31 & 0.000 & 0.277 & 0.301 & 0.301 & 0.292 & \textbf{0.310} \\ 32 & 0.000 & 0.266 & 0.290 & 0.292 & 0.282 & \textbf{0.298} \\ 33 & 0.000 & 0.258 & 0.280 & 0.282 & 0.275 & \textbf{0.286} \\ 34 & 0.000 & 0.248 & 0.271 & 0.271 & 0.266 & \textbf{0.276} \\ 35 & 0.000 & 0.240 & 0.262 & 0.262 & 0.257 & \textbf{0.266} \\ \hline \end{tabular} } \end{table}

\section{Conclusion}\label{sec:conclusion}
Gaussian Boson Sampling (GBS) has demonstrated its potential in quantum supremacy in sampling experiments 
over the past decade, and in this work we extend its exploration toward practical relevance in industrially motivated problems. We consider the densest $k$-subgraph problem (D$k$SP), a central graph optimization task involving the identification of dense regions within a network. To enable a principled evaluation, a comprehensive benchmarking framework for hybrid approaches is devised, comparing GBS-based algorithms against established classical baselines. 

It is found that standard GBS with hard post-selection is inherently inefficient for D$k$SP due to strict cardinality constraints and a sampling bias toward lower photon counts. This motivates the incorporation of lightweight classical refinement techniques to enhance raw GBS outputs. The proposed degree-based, edge-based, and ACO-based post-processing strategies improve solution quality and sampling efficiency, with edge-based and ACO-based refinement showing particularly strong performance. The post-processing methods are particularly advantageous in the presence of realistic photon loss and maintain the performance across random and community-structured graphs to real-world network instances.

To quantify the trade-off between quantum-sample requirements, classical computational cost, and solution quality, an ingenious resource analysis is devised, demonstrating that the improvements obtained through post-processing can be achieved with practical classical refinement costs. Overall, we show that while GBS with post-selection alone is insufficient, its integration with classical post-processing yields a practical and competitive hybrid framework, effectively leveraging the structural bias of GBS toward dense subgraphs.

To further substantiate these conclusions, the proposed framework needs to be experimentally implemented on photonic quantum hardware. Building on the cost analysis framework, further examinations are required to study the scalability of these hybrid approaches and benchmark their performance against state-of-the-art classical methods. 
Finally, a natural direction for future work on the quantum component of our framework is to move beyond fixed encodings and make the photonic sampler itself trainable, for instance through learnable graph-embedding strategies, parameterized state initialization, and adaptive measurement circuits. Such trainable elements would allow the sampling bias of the quantum device to be more closely aligned with the structure of the densest k-subgraph instance, which we expect offers a promising avenue toward realizing practical quantum advantage with GBS as relevant primitive within this hybrid framework.



\bibliographystyle{IEEEtran}
\bibliography{bib}
\end{document}